\documentclass[showpacs,amsmath,amssymb,aps,showkeys,floatfix,prd,a4paper]{revtex4}

\usepackage[dvips]{graphicx}
\usepackage{dcolumn}
\usepackage{bm}
\usepackage{epsfig}
\usepackage{amsfonts}
\usepackage{amssymb,amscd}

\def\lsim{\raise0.3ex\hbox{$<$\kern-0.75em\raise-1.1ex\hbox{$\sim$}}}
\def\gsim{\raise0.3ex\hbox{$>$\kern-0.75em\raise-1.1ex\hbox{$\sim$}}}

\def\beq{\begin{equation}}
\def\eeq{\end{equation}}
\def\bea{\begin{eqnarray}}
\def\eea{\end{eqnarray}}
\def\bq{\begin{quote}}
\def\eq{\end{quote}}

\newcommand{\rr}{\mbox{\boldmath $r$}}

\def\gappeq{\mathrel{\rlap {\raise.5ex\hbox{$>$}}
{\lower.5ex\hbox{$\sim$}}}}

\def\lappeq{\mathrel{\rlap{\raise.5ex\hbox{$<$}}
{\lower.5ex\hbox{$\sim$}}}}

\def\Toprel#1\over#2{\mathrel{\mathop{#2}\limits^{#1}}}

\begin{document}


\title{Probing the Color Glass Condensate  in $pp$ collisions at forward rapidities and very low transverse momenta}

\author{V.~P. Gon\c{c}alves}
\email{barros@ufpel.edu.br}
\affiliation{High and Medium Energy Group, \\
Instituto de F\'{\i}sica e Matem\'atica, Universidade Federal de Pelotas\\
Caixa Postal 354, CEP 96010-900, Pelotas, RS, Brazil}
\author{M. L. L. da Silva}
\affiliation{High and Medium Energy Group, \\
Instituto de F\'{\i}sica e Matem\'atica, Universidade Federal de Pelotas\\
Caixa Postal 354, CEP 96010-900, Pelotas, RS, Brazil}

\date{\today}

\begin{abstract}
The description of the hadron production at very forward rapidities and low transverse momentum is usually made using phenomenological models based on nonperturbative physics. However, at high energies and large rapidities  
the wave function of one of the projectiles is probed at  very small Bjorken $x$, being  characterized by a large number of gluons. In this kinematical regime,  a new state of matter - the Color Glass Condensate (CGC) -  
is expected to be formed. One the main characteristics of such system is the presence of a new dynamical momentum scale, the saturation scale $Q_s$, which can assume values  very larger than the QCD confinement scale $\Lambda_{QCD}$ and  give the scale of the running coupling constant. In this paper we assume that in particular  kinematical region probed by LHC forward (LHCf) experiment the saturation scale can be considered the hard momentum scale present in the process and calculate the forward neutral pion production  at very low-$p_T$ using a perturbative approach. We demonstrate that the CGC formalism is able to successfully describe the LHCf data, which can be considered as a compelling indication of the presence of non-linear QCD effects at LHC energies.

\end{abstract}


\maketitle


\section{ Introduction} 
The Large Hadron Collider (LHC) has opened up a new frontier in high energy hadron - hadron collisions, allowing to test the Quantum Chromodynamics in unexplored regimes of energy, density and rapidities, considering different configurations of the colliding hadrons (protons and nuclei) (For a recent review see e.g. \cite{cgc_review}). In particular, the LHC experiments has unprecedented capacities to study several subjects associated to {\it forward physics}  as, for instance, soft and hard diffraction, exclusive production of new mass states, low-$x$ dynamics and other important topics (For a review see e.g. Ref. \cite{blois}). Furthermore, the study of the forward particle production at LHC is expected to be able to  constrain the model used as input to the modelling of high-energy air showers in cosmic ray experiments.

Forward physics is characterized by the production of particles with relatively small transverse momentum, being traditionally associated with soft particle production, which is intrinsically nonperturbative and not amenable to first-principles analysis.
Very recently, in  Ref. \cite{lhcf}, the inclusive production rate of neutral pions at rapidities larger  than $y = 8.9$ and very low transverse momentum $p_T$ ($\le 0.6$ GeV) were measured by the Large Hadron Collider forward (LHCf) experiment in $pp$ collisions at $\sqrt{s} = 7$ TeV. 
In \cite{lhcf} the transverse momentum spectra were compared with the predictions of several hadronic interaction models based on distinct assumptions. In particular, with models which assume that the particle production at very forward rapidities is dominated by nonperturbative (soft) physics, which is justified, {\it in a first approximation}, considering the range of transverse momentum probed by the experiment. 

In this letter we propose a distinct perspective for the description of the LHCf data. Basically, we take into account that at LHC energies and very forward rapidities, the wave function of one of the projectiles is probed at large Bjorken $x$ and that of the other at very small $x$. The latter  is characterized by a large number of gluons, which is expected to form a new state of matter - the Color Glass Condensate (CGC) -  where the gluon distribution saturates and non-linear coherence phenomena dominate (For a review see e.g. \cite{cgc_review}). Such a system is endowed with a new dynamical momentum scale, the saturation scale $Q_s$, which controls the main characteristic of the particle production and whose evolution  
is described by an infinite hierarchy of coupled equations for the correlators of  Wilson lines \cite{BAL,KOVCHEGOV,CGC}.  
At  large energies and rapidities, $Q_s$ is expected to become  very larger than the QCD confinement scale $\Lambda_{QCD}$ and  give the scale of the running coupling constant. 
Our main assumption is that in the particular  kinematical region probed by LHCf, the saturation scale is very
larger than the QCD confinement scale $\Lambda_{QCD}$ and is the dominant momentum scale present in the process, which implies that $\alpha_s(Q_s^2) \ll 1$ and allows to calculate the neutral pion production at very low-$p_T$ using a perturbative approach. It is important to emphasize that this assumption also is implicitly present in the CGC calculations 
of the bulk features of the RHIC and LHC data such as the energy, rapidity and centrality dependence of particle multiplicities (See, e.g. \cite{review_results_cgc}).

\section{Hadron production at forward rapidities}

The description of  hadron production at large transverse momentum $p_T$ is one the main examples of a hard process in perturbative QCD (pQCD). It  can be accurately described within collinear factorization, by combining partonic cross-sections computed to some fixed order in perturbation theory with parton distribution and fragmentation functions whose evolution is computed by solving the Dokshitzer - Gribov - Lipatov - Altarelli - Parisi (DGLAP) equations \cite{dglap} to the corresponding accuracy in pQCD.  The high transverse momentum $p_T$ of the produced hadron insures applicability of pQCD, which is expected to fail to low-$p_T^2$. Furthermore, at forward rapidities the small-$x$ evolution becomes important, leading to an increasing in the density of gluons and in their transverse momentum. Because of that, in this kinematical range their evolution in transverse momenta cannot be disregarded, which implies that at very forward rapidities the collinear factorization is expected to breakdown. An alternative is the description of the hadron production using the $k_T$-factorization scheme, which is based on the unintegrated gluon distributions whose evolution is described by  the Balitsky-Fadin-Kuraev-Lipatov (BFKL) equation \cite{bfkl}. However, if the transverse momentum of some of the produced particles is comparable with the saturation momentum scale, the partons from one projectile scatter off a dense gluonic system in the other projectile. In this case the  parton undergo multiple scatterings, which cannot be encoded in the traditional (collinear and $k_T$) factorization schemes.  
As pointed in Ref. \cite{difusivo}, the forward hadron production in hadron-hadron collisions is  a typical example of a dilute-dense process, which is an ideal system to study the small-$x$ components of the  target wave function.  In this case the cross section is expressed as a convolution of the standard parton distributions for the dilute projectile, the dipole-hadron scattering amplitude (which includes the high-density effects) and the parton fragmentation functions.  Basically, assuming this generalized dense-dilute factorization, the minimum bias invariant yield for single-inclusive hadron production in hadron-hadron processes is described in the CGC formalism  by \footnote{The Eq. (\ref{eq:final}) was improved in Ref. \cite{kovner_ine} by the inclusion of inelastic contributions which are important at high transverse momentum ($p_T \gg Q_s$). However, in the kinematical range of interest in this paper (very small-$p_T$, with $p_T \lesssim Q_s$) this new contribution is negligible.}
 \cite{dhj}
\begin{widetext}
\begin{eqnarray}\label{eq:hadron}
 \frac{d^2N^{pp\rightarrow \pi^0X}} {dyd^2p_T}&=& {\cal{K}} \frac{1}{(2\pi)^2}
\int_{x_F}^1 dx_1 \frac{x_1}{x_F}\left[ f_{q/p}(x_1,\mu^2){\cal{N}_F}\left(x_2,\frac{x_1}{x_F}p_T\right)
D_{\pi^0/q}\left(\frac{x_F}{x_1},\mu^2\right)
\right.\nonumber\\
&+& \left.   f_{g/p}(x_1,\mu^2){\cal{N}_A}\left(x_2,\frac{x_1}{x_F}p_T\right)D_{\pi^0/g}\left(\frac{x_F}{x_1},\mu^2\right)  \right]\,\,,
\label{eq:final}
\end{eqnarray}
\end{widetext}
where $p_T$, $y$ and $x_F$ are the transverse momentum, rapidity and the Feynman-$x$
of the produced hadron, respectively. The $\cal{K}$-factor mimics the effect of higher order corrections and, effectively, of other dynamical effects not included in the CGC formulation.  The variable $x_1$ denotes the momentum
fraction of a projectile parton,    $f(x_1,\mu^2)$ the projectile parton
distribution functions  and $D(z, \mu^2)$ the parton fragmentation
functions into neutral pions. These quantities  evolve according to the 
DGLAP evolution equations \cite{dglap} and obey the momentum
sum-rule. It is useful to assume $\mu^2 = p_T^2$ (See discussion below).  Moreover, $x_F=\frac{p_T}{\sqrt{s}}e^{y}$ and the momentum fraction of the target partons is given by $x_2=x_1e^{-2y}$ (For details see e.g. \cite{dhj}).
 In Eq. (\ref{eq:final}), ${\cal{N}_F}(x,k)$  and  ${\cal{N}_A} (x,k)$ are the fundamental and adjoint representations of the forward dipole amplitude in momentum space and are given by
\begin{eqnarray}
{\cal{N}_{A,F}}(x,p_T)=  \int d^2 r e^{i\vec{p_T}\cdot \vec{r}}\left[1-{\cal{N}_{A,F}}(x,r)\right]\,\,,
\end{eqnarray}
where  ${\cal{N}_{A,F}}(x,r)$ 
encodes all the
information about the hadronic scattering, and thus about the
non-linear and quantum effects in the hadron wave function.
In the large-$N_c$ limit we have the following relation between the
adjoint and fundamental representations:
\begin{eqnarray}
{\cal{N}_{A}}(x,r) = 2 {\cal{N}_{F}}(x,r) - {\cal{N}}^2_{\cal{F}}(x,r)\,\,.
\end{eqnarray}
The scattering amplitude ${\cal{N}_A}(x,r)$ 
can be obtained by solving the BK evolution equation \cite{BAL,KOVCHEGOV} or considering phenomenological QCD inspired models to describe the interaction of the dipole with the target. BK equation is the simplest nonlinear evolution equation
for the dipole-hadron scattering amplitude, being 
actually a mean field version
of the first equation of the B-JIMWLK hierarchy \cite{BAL,CGC}. At 
order (LO), and in the translational invariance approximation---in which the scattering
amplitude does not depend on the collision impact parameter $\bm{b}$--- the BK equation reads
	\begin{equation}\label{eq:bklo}
		\frac{\partial {\cal{N}_A}(r,Y)}{\partial Y} = \int {\rm d}\bm{r_1}\, K^{\rm{LO}}
		(\bm{r,r_1,r_2})
		[{\cal{N}_A}(r_1,Y)+{\cal{N}_A}(r_2,Y)-{\cal{N}_A}(r,Y)-{\cal{N}_A}(r_1,Y){\cal{N}_A}(r_2,Y)],
	\end{equation}
where  $Y\equiv \ln(x_0/x)$ ($x_0$ is the value of $x$ where the evolution starts),  $\bm{r_2 = r-r_1}$ and $K^{\rm{LO}}$ is the evolution kernel, given by
	\begin{equation}\label{eq:klo}
		K^{\rm{LO}}(\bm{r,r_1,r_2}) = \frac{N_c\alpha_s}{2\pi^2}\frac{r^2}{r_1^2r_2^2},
	\end{equation}
where $\alpha_s$ is the (fixed) strong coupling constant. 
In its linear
version, the BK equation corresponds to the BFKL equation
\cite{bfkl}.
The solution of the LO BK equation implies that the saturation scale grows much faster with increasing energy
($Q_s^2\sim x^{-\lambda}$, with $\lambda \approx 0.5$) than that
extracted from phenomenology ($\lambda \sim 0.2-0.3$).

In the last years the next-to-leading order corrections to the  BK equation were
 calculated  \cite{kovwei1,javier_kov,balnlo} through the ressumation of $\alpha_s N_f$ contributions to 
all orders, where $N_f$ is the number of flavors. 
The  improved  BK equation is given in terms of a  
running coupling and a subtraction term, with the latter accounting for conformal, non running coupling contributions. In the prescription proposed by Balitsky in \cite{balnlo} to single out the ultra-violet divergent contributions from the finite ones that originate after the resummation of quark loops, the contribution of the subtraction term is minimized at large energies. In \cite{rcbk} this contribution was disregarded, and the improved BK equation was numerically solved replacing the leading order kernel  in Eq. (\ref{eq:bklo}) by the modified kernel which includes the running coupling
corrections and  is given by \cite{balnlo}
	\begin{equation}\label{eq:krun}
		K^{\rm{Bal}}(\bm{r,r_1,r_2})=\frac{N_c\alpha_s(r^2)}{2\pi^2}
		\left[\frac{r^2}{r_1^2r_2^2} + \frac{1}{r_1^2}\left(\frac{\alpha_s(r_1^2)}
		{\alpha_s(r_2^2)}-1\right)+\frac{1}{r_2^2}\left(\frac{\alpha_s(r_2^2)}
		{\alpha_s(r_1^2)}-1\right)\right].
	\end{equation}
 The solution of the improved BK equation was studied in detail in Ref. 
\cite{javier_kov}. The running of the coupling reduces 
the speed of the evolution to values compatible with experimental data, with the geometric 
scaling regime being reached only at ultra-high energies. In \cite{rcbk} a global 
analysis of the small $x$ data for the proton structure function using the improved BK 
equation was performed  (See also Ref. \cite{weigert}). In contrast to the  BK  equation 
at leading logarithmic $\alpha_s \ln (1/x)$ approximation, which  fails to describe the 
HERA data, the inclusion of running coupling effects in the evolution renders the BK equation 
compatible with them (See also \cite{vic_joao,alba_marquet,vic_anelise}).
 In what follows we 
consider the BK predictions for ${\cal{N}}(x,\rr)$ (from now on called rcBK) obtained using the McLerran-Venugopalan model for the  initial 
condition \cite{MV}, which is given by
	\begin{equation}
		{\cal{N}}^{\rm{MV}}(r,Y=0)=1-\exp\left[-\left(\frac{r^2Q_{s0}^2}{4}
		\right)^\gamma\ln\left(\frac{1}{r\Lambda_{\rm{QCD}}}+e\right) \right],
	\end{equation}
where $Q_{s0}^2$ is the initial saturation scale squared and $\gamma$ is an anomalous dimension. Both parameters are obtained from the fit to $F_2$ data and are given by 
$Q_{s0}^2=0.15$ GeV$^2$, $\gamma = 1.13$. Moreover, $\Lambda_{\rm{QCD}} = 0.241$ GeV and $x_0 = 0.01$. It is important to emphasize that the solutions of the improved BK equation has been obtained considering that the running coupling is evaluated according to the usual one-loop QCD expression
\begin{eqnarray}
\alpha_s(r^2) = \frac{12 \pi}{(11 N_c - 2 N_f)\frac{4C^2}{r^2\Lambda^2_{\rm{QCD}}}}
\label{alfa}
\end{eqnarray}
at dipoles of small size ($r < r_f$) and frozen at larger sizes ($r>r_f$) to the fixed value $\alpha_s(r_f) = 0.7$. 
The factor $C^2$ in Eq. (\ref{alfa}) is a free parameter which is fixed by the data, being equal to 6.5.

For comparison we also consider the  phenomenological model proposed in \cite{buw} (denoted BUW model hereafter),  which parametrize the adjoint dipole scattering amplitude as follows
\begin{eqnarray}
{\cal{N}_A}(x,p_T)= - \int d^2 r e^{i\vec{p_T}\cdot \vec{r}}\left[1-\exp\left(-\frac{1}{4}(r^2Q_s^2(x))^{\gamma(p_T,x)}\right)\right]\,\,,
\end{eqnarray}
where $\gamma$ is assumed to be a function of $p_T$, rather than $r$, in order to make easier the evaluation of its Fourier transform, and is given by  $\gamma(p_T,x)=\gamma_s+\Delta\gamma(p_T,x)$, where  $\gamma_s = 0.628$ and  \cite{buw}
\begin{eqnarray}
\label{BUWeq}
\Delta \gamma(p_T,x) = \Delta \gamma_{BUW} =(1-\gamma_s)\frac{(\omega^a-1)}{(\omega^a-1)+b}.
\end{eqnarray}
In the expression above, $\omega  \equiv p_T/Q_s(x)$ and the two free parameters $a=2.82$ and $b=168$ are fitted in such a way to describe the RHIC data on hadron production. It is clear, from Eq.(\ref{BUWeq}), that this model satisfies the property of geometric scaling \cite{scaling,marquet,prl,prl1}, since $\Delta\gamma$ depends on $x$ and $p_T$ only through the variable $p_T/Q_s(x)$.
 Besides, in comparison with other phenomenological parameterizations, in the BUW model, the large $p_T$ limit, $\gamma \rightarrow 1$, is approached much faster, which implies different predictions for the large $p_T$ slope of the hadron and photon yield (For a detailed discussion see Ref. \cite{marcos_vic}).

\section{Results} 
In what follows we will use Eq. (\ref{eq:final}), which is based on the CGC formalism, to calculate the neutral pion production in $pp$ collisions at very forward rapidities.  This formalism has successfully been applied to single and double inclusive particle production in proton - proton and proton - nucleus collisions at high energy. 
In particular, the description of the observed suppression of the normalized hadron production transverse momentum in $dAu$ collisions as compared to $pp$ collisions has been considered an important signature of the Color Glass Condensate physics (See e.g. \cite{cgc_review, review_results_cgc}). Previous studies of the hadron production at forward rapidities  applied the  CGC formalism for the kinematical region where $p_T\ge 1$ GeV and assumed that  the hadron transverse momentum provide the hard scale for the parton distribution and fragmentation functions \cite{dhj,buw,marcos_vic,alba_marquet,jamal_amir,Goncalves:2006yt}. However, the latter assumption is not valid in the kinematical region probed by LHCf. On the other hand, one the main consequences of CGC formalism is that a system with a high gluonic density is  endowed with a new dynamical momentum scale, the saturation scale $Q_s$, which controls the main characteristic of the particle production. At the LHC, $Q_s$ is expected to be in range 2-5 GeV depending upon the total energy, the rapidity of the produced particles, and the nature of the hadron (proton or lead nucleus).
In particular, we expect that at  large  rapidities, $Q_s$ becomes  very larger than the QCD confinement scale $\Lambda_{QCD}$ and give the scale of the running coupling constant, making it small $\alpha_s(Q_s^2) \ll 1$. It allows to extend our ability to calculate particle production at very low $p_T$ in a small-coupling framework. Our basic assumption will be that gluon saturation turns particle production into a one scale problem, with $Q_s$ as the only scale apart from the size of the system. As consequence we will assume in Eq. (\ref{eq:final}) that the factorization scale $\mu$ is equal to the saturation scale.

In Fig. \ref{fig1} we present our predictions for the production of neutral pions in $pp$ collisions at forward rapidities and compare our results with the LHCf data \cite{lhcf}. Our predictions were obtained for the central value of the distinct rapidity range. 
In our calculations we use the CTEQ5 parametrization \cite{cteq}  for the parton distribution functions  and the KKP parametrization for the  fragmentation functions \cite{kkp}.   Moreover, we consider the rcBK and BUW models for the scattering amplitude. 
As emphasized above, in our calculations we will assume, in a first approximation, that $\mu^2 = Q_s^2$, with $Q_s^2 = Q_{0s}^2 (\frac{x_0}{x_2})^\lambda$. 
In the  kinematical region probed at LHCf  the values of $x_2$ are  $\lesssim 10^{-9}$.
Following \cite{dhj} we take $x_0 = 10^{-4}$ and $\lambda = 0.3$. 
 Moreover, as in Ref. \cite{jamal_amir} we assume that $Q_{0s}^2 = 0.168$ GeV$^2$. However, as already pointed out in \cite{jamal_amir}, smaller values may be more preferable, especially at very forward rapidities. In order to verify this result, in our analysis we also consider two other possibilities: $Q_{0s}^2 = 0.1$ GeV$^2$ and  $Q_{0s}^2 = 0.05$ GeV$^2$.
As in previous calculations \cite{dhj,buw,marcos_vic,alba_marquet,jamal_amir,Goncalves:2006yt} there is only one free parameter in our calculation: the $\cal{K}$-factor. It is determined in such way to provide the best description of the experimental data. We verified that an identical $\cal{K}$-factor ($\cal{K}$ = 7) for all rapidities alllows to describe the data, independent of the model used for the scattering amplitude. 
Moreover, the rcBK and BUW models for the scattering amplitude predict almost identical
momenta spectra for  $Q_{0s}^2= 0.168$ GeV$^2$. A similar result is obtained for the other values of $Q_{0s}^2$. This result is directly associated to the similar behaviour of the dipole scattering amplitude at large pair separations predicted by these two models (See Fig. 1 in Ref. \cite{marcos_vic}), which is the region probed in $pp$ collisions at forward rapidities.
Assuming  smaller values of $Q_{0s}^2$ we obtain similar predictions for the momenta spectra, with a mild difference at larger values of $p_T$.

Finally, we analyze the dependence of our predictions on the choices for the fragmentation functions $D_{\pi^0/i}$, parton distributions $ f_{i/p}$ and factorization scale $\mu$. We have checked that the fragmentation functions proposed in Ref. \cite{hkns} are also consistent with our results. Moreover, our predictions are almost identical if the CTEQ6 parametrization \cite{cteq6}  for the parton distribution functions are used in the calculations. However, as already pointed out in \cite{alba_dumitru}, the   value of the ${\cal{K}}$ - factor necessary in order to describe the data is strongly dependent on the choice of the factorization scale $\mu$.  In particular, the value of the ${\cal{K}}$ - factor is reduced for ${\cal{K}} = 3$ if we assume that $\mu = Q_s/2$. In Fig.  \ref{fig2} we compare the resulting  predictions with those obtained using $\mu = Q_s$ and  ${\cal{K}} = 7$.  We have that both choices allow us to describe the LHCf data. It is important to emphasize that the value  ${\cal{K}} = 3$ is  similar to the values used in Ref. \cite{alba_dumitru} to describe the CMS $pp$ data.
Our  main conclusion is that our predictions are almost independent of the following choices: scattering amplitude, parton distribution and fragmentation functions and the initial saturation scale $Q_{0s}$. As we can see from Figs. \ref{fig1} and \ref{fig2}, the CGC formalism is able to successfully describe the LHCf data if we assume the emergence of the saturation scale as the hard scale of the problem. Only at very low $p_T$ ($< 0.1$ GeV) our predictions behave harder than the data, which can be an indication of soft contributions. However, at $p_T \ge 0.1$ the experimental data are described quite well assuming the dilute - dense factorization expressed in Eq. (\ref{eq:final}). We believe that this results is a compelling indication of the presence of non-linear QCD effects.

\section{ Summary} 
At very high energies  the traditional separation between hard and soft QCD dynamics can be oversimplified due to the presence of novel semihard scales generated dinamically, which allows to understand highly nonperturbative phenomena in QCD by using weak coupling methods.  In this paper we assumed that in the particular kinematical region probed by LHCf the intrinsic momentum scale $Q_s$, associated to the gluon saturation is much larger than $\Lambda_{QCD}$ and, consequently,  that  $\alpha_s(Q_s^2) \ll 1$, which implies that the neutral pion production can be calculated using a perturbative approach which includes  the large gluon density present in the hadron target. Basically, we assumed  the emergence of saturation scale as a hard momentum scale at very forward rapidities and  extended at very low $p_T$ the dilute - dense factorization, derived in the CGC formalism.   
Our results demonstrate that  for LHC energies and very forward rapidities the neutral pion production at low transverse momentum measured by LHCf collaboration can be quite well described considering the CGC formalism. We believe that this result can be considered a signature of the presence of non-linear QCD effects at high energies probed at LHC. Furthermore, it can be considered as a indication that the gluon saturation effects cannot be disregarded  
for the modelling of the high-energy air showers.

\begin{widetext}

\begin{figure}[t]
\centerline{\psfig{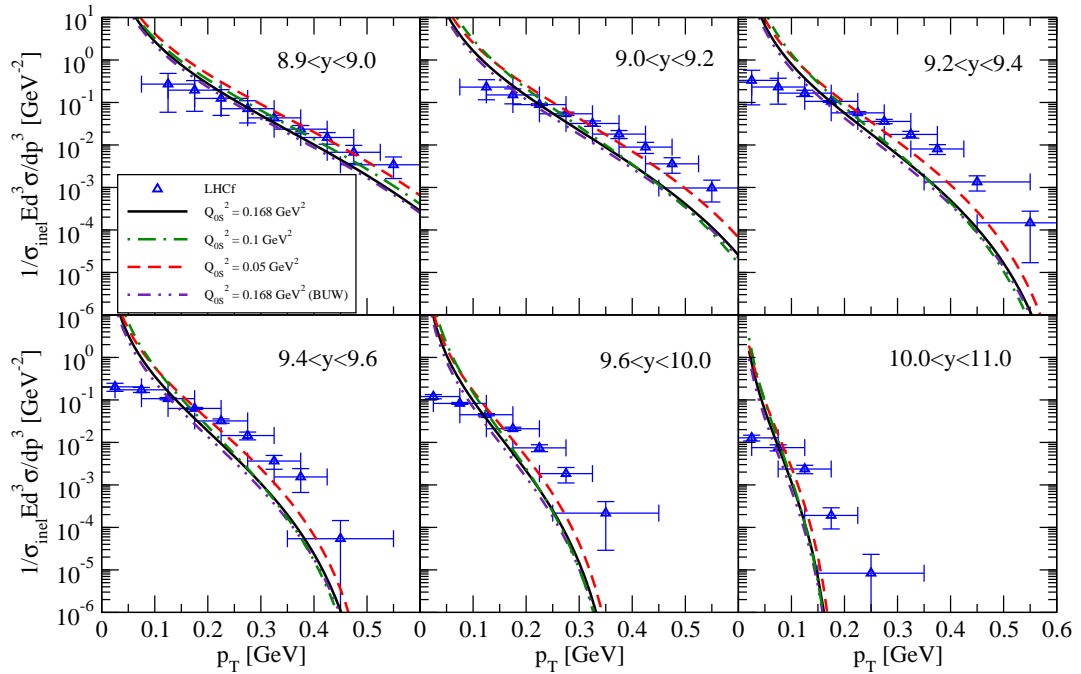}}
 \caption{Neutral pion yields in proton - proton collisions at $\sqrt{s} = 7$ TeV and different values of the rapidity range (See text). Data by the LHCf collaboration \cite{lhcf}.}
\label{fig1}
\end{figure}

\begin{figure}[t]
\centerline{\psfig{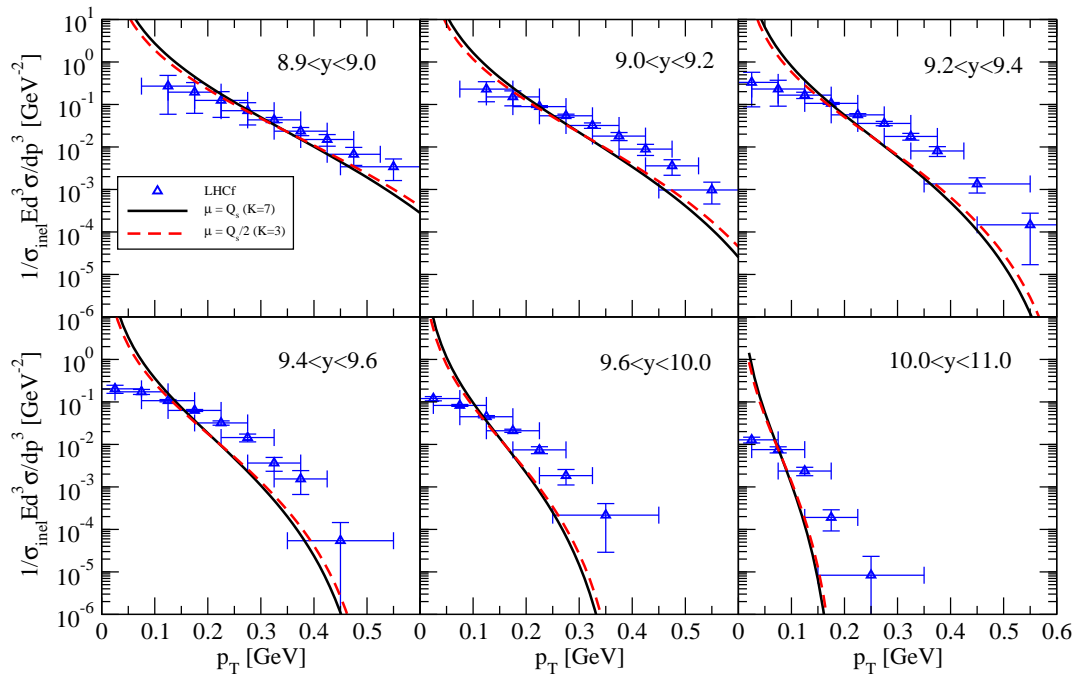}}
 \caption{Dependence of the predictions for neutral pion yields in proton - proton collisions at $\sqrt{s} = 7$ TeV on the factorization scale and $\cal{K}$-factor. Data by the LHCf collaboration \cite{lhcf}.}
\label{fig2}
\end{figure}

\end{widetext}


\section*{Acknowledgements}
 This work was partially financed by the Brazilian funding agencies CNPq and FAPERGS.



\end{document}